# Multiple membership multilevel models

George Leckie

Centre for Multilevel Modelling and School of Education, University of Bristol

**Address for correspondence**

Centre for Multilevel Modelling

School of Education

University of Bristol

35 Berkeley Square

Bristol

BS8 1JA

United Kingdom

g.leckie@bristol.ac.uk

# Multiple membership multilevel models

**Summary**

Multiple membership multilevel models are an extension of standard multilevel models for non-hierarchical data that have multiple membership structures. Traditional multilevel models involve hierarchical data structures whereby lower-level units such as students are nested within higher-level units such as schools and where these higher-level units may in turn be nested within further groupings or clusters such as school districts, regions, and countries. With hierarchical data structures, there is an exact nesting of each lower-level unit in one and only one higher-level unit. For example, each student attends one school, each school is located within one school district, and so on. However, social reality is more complicated than this, and so social and behavioural data often do not follow pure or strict hierarchies. Two types of non-hierarchical data structures which often appear in practice are cross-classified and multiple membership structures. In this article, we describe multiple membership data structures and multiple membership models which can be used to analyse them.

**Introduction**

In multiple membership data structures, there is not an exact nesting of each lower-level unit in one and only one higher-level unit. Rather, lower-level units are nested within multiple higher-level units. An example in educational research arises in studies of student attainment where students are often taught by more than one teacher and so students are described as being multiple members of those teachers. An example in health services research arises in studies of hospital patient outcomes. Hospital patients will often be seen and treated by not one doctor, but multiple doctors, and so here patients are multiple members of doctors.

**The consequences of ignoring multiple membership structures**

It is important to incorporate multiple membership structures into our models when they arise in the data. In our educational research example, it seems intuitive that we should recognize the contribution or effect that every teacher has on student attainment. Similarly, in our health services research example, it seems intuitive to recognize the role that every doctor plays in contributing to patients' outcomes. Ignoring multiple membership structures by, for example, simply assigning each lower-level unit to just one of their higher-level units has serious consequences as this will underestimate the degree of higher-level clustering and therefore the importance of the higher-level units on the outcome variable (Chung and Beretvas, 2012). Thus, ignoring the multiple membership structures in our examples would lead us to underestimate the importance of teachers as a determinant of student attainment and of doctors as a determinant of patient outcomes.

**Multiple membership weights**

An important feature of multiple membership data structures is that the degree to which each lower-level unit belongs to each higher-level unit will often vary across these higher-level units. Multiple membership weights are used to quantify this. In our educational research example, students may spend more time with some teachers than others. Here we would define multiple membership weights as the proportion of time spent with each teacher. Thus, if a student is taught for two lessons a week by teacher A and three lessons a week by teacher B, we would assign multiple membership weights to that student of 0.4 and 0.6 for teachers A and B, respectively. These weights reflect the fact that we would expect teacher B to be more influential in determining the student's outcome than teacher A. Other weighting schemes may equally be applied, and it can be interesting to explore a range of weighting schemes as part of a sensitivity analysis for the model.

**Model equations**

The multiple membership model for the above educational research example, where we adjust for a single covariate, can be written using "classification notation" (Browne et al., 2001) as

$$y_i = \beta_0 + \beta_1 x_i + \sum_{j \in \text{teacher}(i)} w_{j,i} u_j + e_i$$

$$u_j \sim \text{N}(0, \sigma_u^2)$$

$$e_i \sim \text{N}(0, \sigma_e^2)$$

where $y_i$ denotes the attainment of student $i$, $\beta_0$ is the model intercept, $x_i$ denotes the value of the covariate for that student, $\beta_1$ is the associated slope coefficient, $\sum_{j \in \text{teacher}(i)} w_{j,i} u_j$ is a weighted sum of teacher effects where the multiple membership weight $w_{j,i}$ measures the extent to which student $i$ belongs to teacher $j$ with associated effect $u_j$, and $e_i$ is the residual error term. The term $\text{teacher}(i)$ is a 'classification function' which returns the subset of teachers who taught student $i$. The teacher random effects and residual errors are assumed normally distributed with zero means and constant variances where $\sigma_u^2$ denotes the between-teacher variance and $\sigma_e^2$ denotes the student-level residual error variance. The magnitudes of the variance components may then be compared to make statements about the relative contribution of each classification to the variation in the response, having adjusted for the covariate.

When higher-level explanatory variables are entered into multiple membership models, these variables should be entered in such a way that they respect the multiple membership structure of the data. In our educational research example, suppose that we want to estimate the effect of teacher gender on student attainment. When students are taught by multiple teachers, some teachers will be male, others female. In a multiple membership model, we must recognize the gender of every teacher that has taught each student. To do this, we would derive, and enter into the model specification, a new teacher-level variable which is the weighted average of the teacher gender dummy variable across the series of teachers that have taught each student.

**Estimation and software**

Multiple membership models can be estimated by both frequentist (e.g., maximum likelihood) and Bayesian (e.g., Markov chain Monte Carlo, MCMC) methods. However, few multilevel modelling software packages provide specific routines for fitting such models. A notable exception is the MLwiN (Charlton et al., 2019) software which can fit multiple membership models by both frequentist methods (Rasbash et al., 2012, Chapter 19) and Bayesian methods (Browne, 2019, Chapter 16). For more complex models, for example, with discrete responses or many different classifications, Bayesian estimation will often be considerably more computationally efficient than frequentist estimation. MLwiN can be run from within both the R and Stata software (Leckie and Charlton, 2013; Zhang et al., 2016).

**Modelling extensions**

The social and behavioural data that arise from social reality will often have far more complex data structures than those given in the educational and health services research examples above. To realistically model this complexity, we must often combine hierarchical, cross-classified, and multiple membership data structures in our multilevel models. In our health services research example, patients were multiple members of doctors within a single hospital. When our data come from many hospitals, we would account for hospital level clustering by building in the hierarchical structure of doctors nested within hospitals into our models. Furthermore, were we to also observe the general practitioners (GPs, i.e., family doctors) who refer patients to their hospitals, we might also want to account for GP level clustering. Here we note that GPs and hospitals are likely to be cross-classified as GPs will tend to refer patients to different hospitals depending on their needs, and each hospital will receive patients from many GPs. We would therefore account for GP level clustering by building in a cross-classification of GPs with hospitals into our models.

An important area of multilevel analysis which combines hierarchical, cross-classified, and multiple membership data structures in the same model occurs in the multilevel analysis of spatial data (Lawson, et al., 2003). For example, in a study of neighbourhood effects on quality of life, we might start by fitting a two-level hierarchical model of individuals nested within neighbourhoods. However, intuitively and from theory, we might expect individuals' quality of life to be influenced by not only their own neighbourhood of residence but also by other nearby neighbourhoods. We can attempt to model this spatial correlation in our models by incorporating a multiple membership structure of individuals belonging to the series of neighbourhoods which border each individual's own neighbourhood. In doing so, we implicitly introduce a cross-classification between the neighbourhood of residence classification and the classification for nearby neighbourhoods. For the multiple membership weights, we would specify weights that are proportional to the length of the shared border of each adjacent neighbourhood or the population of each adjacent neighbourhood.

The last two examples have shown how complex multilevel models can become when we try to extend them to realistically reflect the complex data structures that arise in social reality. Unit diagrams and classification diagrams have both been proposed as helpful aides to understanding and communicating complex multilevel data structures (Browne et al., 2001). Similarly, classification notation, which avoids the proliferation of subscripts that arises when we combine many different data structures in a single model, has been proposed as an alternative to standard notation when expressing these models in equation form (Browne et al., 2001).

An interesting and clever use of multiple membership models is to handle missing identifiers in hierarchical data (Hill and Goldstein, 1998). For example, in a school effectiveness study, we might know that each student attends only one school, but for some students, we do not observe which school that is. If however we do observe where these students live, then we can specify the probability that each student attends each of several local schools, perhaps as an inverse function of distance. These probabilities are entered into the model as multiple membership weights.

**Further reading**

Introductory, intermediate, and advanced treatments of multilevel models are given in the multilevel modelling textbooks by Snijders and Bosker (2012), Raudenbush and Bryk (2002), and Goldstein (2011), respectively. Accessible introductions to multiple membership models are given by the report by Fielding and Goldstein (2006) and the book chapter by Beretvas (2010). More advanced treatments of multiple membership models are provided in the multilevel textbook by Goldstein (2011, Chapter 13), the book chapters on multiple membership models by Rasbash and Browne (2001, 2008), the paper by Browne et al. (2001), and the report by Leckie (2013). An application of multiple membership models to modelling student attainment is provided by Leckie (2009) who model the separate multiple memberships of students within schools and students within neighbourhoods in order to account for student mobility over time. A second application of multiple membership models in this context is by Timmermans et al. (2012) who focus on the separate multiple memberships of students within secondary schools and students within primary schools, again in order to account for student mobility over time.

**References**

Beretvas, S. N. (2010). Cross-classified and multiple membership models. In J. J. Hox & J. K. Roberts (Eds.), Handbook of advanced multilevel analysis. London: Psychology Press.

Browne, W. J. (2019). *MCMC estimation in MLwiN, v3.04*. Centre for Multilevel Modelling, University of Bristol.

Browne, W. J., Goldstein, H., & Rasbash, J. (2001). Multiple membership multiple classification (MMMC) models. *Statistical Modelling*, 1, 103–124.

Charlton, C., Rasbash, J., Browne, W.J., Healy, M. and Cameron, B. (2019) *MLwiN Version 3.04*. Centre for Multilevel Modelling, University of Bristol.

Chung, H., & Beretvas, S. N. (2012). The impact of ignoring multiple membership data structures in multilevel models. *British Journal of Mathematical and Statistical Psychology*, 65, 185–200.

Fielding, A., & Goldstein, H. (2006). Cross-classified and multiple membership structures in multilevel models: An introduction and review. Research report RR791. London: Department for Education and Skills. URL: http://www/bristol.ac.uk/cmm/team/cross-classified-review.pdf.

Goldstein, H. (2011). *Multilevel statistical models* (4th ed.). London: Wiley.

Hill, P. W., & Goldstein, H. (1998). Multilevel modelling of educational data with cross-classification and missing identification of units. *Journal of Educational and Behavioral Statistics*, 23, 117–128.

Lawson, A. B., Browne, W. J., & Vidal Rodeiro, C. L. (2003). *Disease mapping using WinBUGS and MLwiN*. London: Wiley.

Leckie, G. (2009). The complexity of school and neighbourhood effects and movements of pupils on school differences in models of educational achievement. *Journal of the Royal Statistical Society: Series A*, 172, 537–554.

Leckie, G. (2013). Multiple membership multilevel models - concepts. LEMMA VLE Module, 13, 1–61. URL: http://www.bristol.ac.uk/cmm/learning/course.html.

Leckie, G., & Charlton, C. (2013). runmlwin: A Program to Run the MLwiN Multilevel Modelling Software from within Stata. *Journal of Statistical Software*, 52, 1-40.

Rasbash, J., & Browne, W. (2001). Modelling non-hierarchical structures. In A. H. Leyland & H. Goldstein (Eds.), Multilevel modelling of health statistics. Chichester: Wiley.

Rasbash, J., & Browne, W. J. (2008). Non-hierarchical multilevel models. In J. De Leeuw & E. Meijer (Eds.), Handbook of multilevel analysis. New York: Springer.

Rasbash, J., Steele, F., Browne, W. J., & Goldstein, H. (2019). *A user's guide to MLwiN, v3.03*. Centre for Multilevel Modelling, University of Bristol.

Raudenbush, S. W., & Bryk, A. S. (2002*). Hierarchical linear models: Applications and data analysis methods* (2nd ed.). Thousand Oaks, CA: Sage Publications.

Snijders, T. A. B., & Bosker, R. J. (2012*). Multilevel analysis: An introduction to basic and advanced multilevel modeling* (2nd ed.). London: Sage.

Timmermans, A. C., Snijders, T. A. B., & Bosker, R. J. (2012). In search of value added in the case of complex school effects. *Educational And Psychological Measurement*, 73, 210-228.

Zhang, Z., Parker, R., Charlton, C., Leckie, G., & Browne, W. J. (2016). R2MLwiN - A program to run the MLwiN multilevel modelling software from within R. *Journal of Statistical Software,* 72, 10, 1-43. DOI: 10.18637/jss.v072.i10.